\documentclass[10pt]{article}

\usepackage[margin=1in]{geometry}

\usepackage{subcaption}
\usepackage{tikz}
\usetikzlibrary{fit,shapes, arrows, shadows, calc, automata, positioning,matrix,backgrounds,decorations.pathmorphing,backgrounds,petri,decorations.markings,decorations.pathreplacing}
\usepackage{graphicx}
\usepackage{amsfonts}
\usepackage{amsmath}
\usepackage{listings}
\usepackage{bm}
\usepackage{url}

\newcommand{\nop}[1]{}

\newcommand{\R}{\mathcal{R}}
\newcommand{\Rplus}{+}
\newcommand{\Rprod}{*}
\newcommand{\Rzero}{\bm{0}}
\newcommand{\Rone}{\bm{1}}

\title{F-IVM: Learning over Fast-Evolving Relational Data}

\author{
Milos Nikolic$^1$, 
Haozhe Zhang$^2$ 
Ahmet Kara$^2$, 
Dan Olteanu$^2$, 
\\ \\
$^1$University of Edinburgh
\enspace\enspace 
$^2$University of Oxford  
}

\date{}
\begin{document}

\maketitle
\begin{abstract}
  F-IVM is a system for real-time analytics such as machine learning applications over training datasets defined by queries over fast-evolving relational databases. 
  We will demonstrate F-IVM for three such applications: model selection, Chow-Liu trees, and ridge linear regression.
\end{abstract}

\paragraph{Acknowledgements}
This project has received funding from the European Union's Horizon 2020 research and innovation programme under grant agreement No 682588.

\maketitle
\section{Learning Under Updates}

\nop{
F-IVM is an open-source\footnote{\url{https://github.com/fdbresearch/FIVM}} system for real-time analytics over fast-evolving relational databases~\cite{Nikolic:SIGMOD:18}. It is a prime example of how  database technology can help significantly speed up machine learning workloads over relational data. 
}

F-IVM (\url{https://github.com/fdbresearch/FIVM}) is a system for real-time analytics over fast-evolving relational databases~\cite{Nikolic:SIGMOD:18}.

F-IVM innovates on two fronts. 
First, F-IVM puts forward a novel incremental maintenance mechanism for batches of aggregates over arbitrary project-join queries. It constructs a tree of views, with the input relations as leaves, the query as root, and each view defined by the join of its children possibly followed by projecting away attributes. For updates to a relation, it maintains the views along the path to the tree root using delta processing and view materialization. 

Second, F-IVM captures the data-intensive computation of many applications using application-specific rings, which define sum and product operations over data values. The ring of integers suffices to treat updates uniformly for sum-product aggregates over joins: negative/positive tuple multiplicities denote deletes/inserts~\cite{DBT:VLDBJ:2014}. More complex applications call for richer rings or even composition of rings. F-IVM introduces the {\em degree-m matrix} ring to maintain the gradients for linear regression models. Moreover, it uses the same view tree to maintain factorized conjunctive query evaluation, matrix chain multiplication, and linear regression, with the only computational change captured by the ring.

F-IVM differs from existing online learning algorithms in at least two ways. (1) Whereas the latter only consider inserts~\cite{murphy2013}, F-IVM also considers deletes. (2) F-IVM avoids the materialization of the training dataset defined by a feature extraction query over multi-relational data. It casts the data-intensive computation of the learning task as ring computation inside the views and pushes it past the joins and down the view tree of the query. This comes with great performance benefits: F-IVM can maintain model gradients over a join faster than maintaining the join, since the latter may be much larger and have many repeating values. It is also competitive against state-of-the-art incremental view maintenance systems: Experiments showed several orders of magnitude performance speedup over DBToaster~\cite{DBT:VLDBJ:2014}, with an average throughput of 10K updates per second for batches of up to thousands of aggregates over joins of five relations on one thread of a standard Azure machine~\cite{Nikolic:SIGMOD:18}.

We will demonstrate F-IVM's unique ability to maintain the pairwise mutual information (MI) and the covariance matrices (COVAR) over categorical and continuous attributes. These matrices represent the data-intensive computation of common machine learning applications. MI is used for model selection, Chow-Liu trees (optimal tree-shaped Bayesian networks), as cost function in learning decision trees, and in determining the similarity of clusterings of a dataset.  COVAR is used for ridge linear regression~\cite{SOC:SIGMOD:16}, forward/backward model selection, polynomial regression, and factorization machines. 
For this demonstration, we will use {\it model selection, ridge linear regression, and Chow-Liu trees}. Our web-based user interface  uses F-IVM to maintain these applications under updates to the Retailer~\cite{SOC:SIGMOD:16} and Favorita~\cite{favorita} databases.

\begin{figure*}[t]
\includegraphics[width=1.0\linewidth]{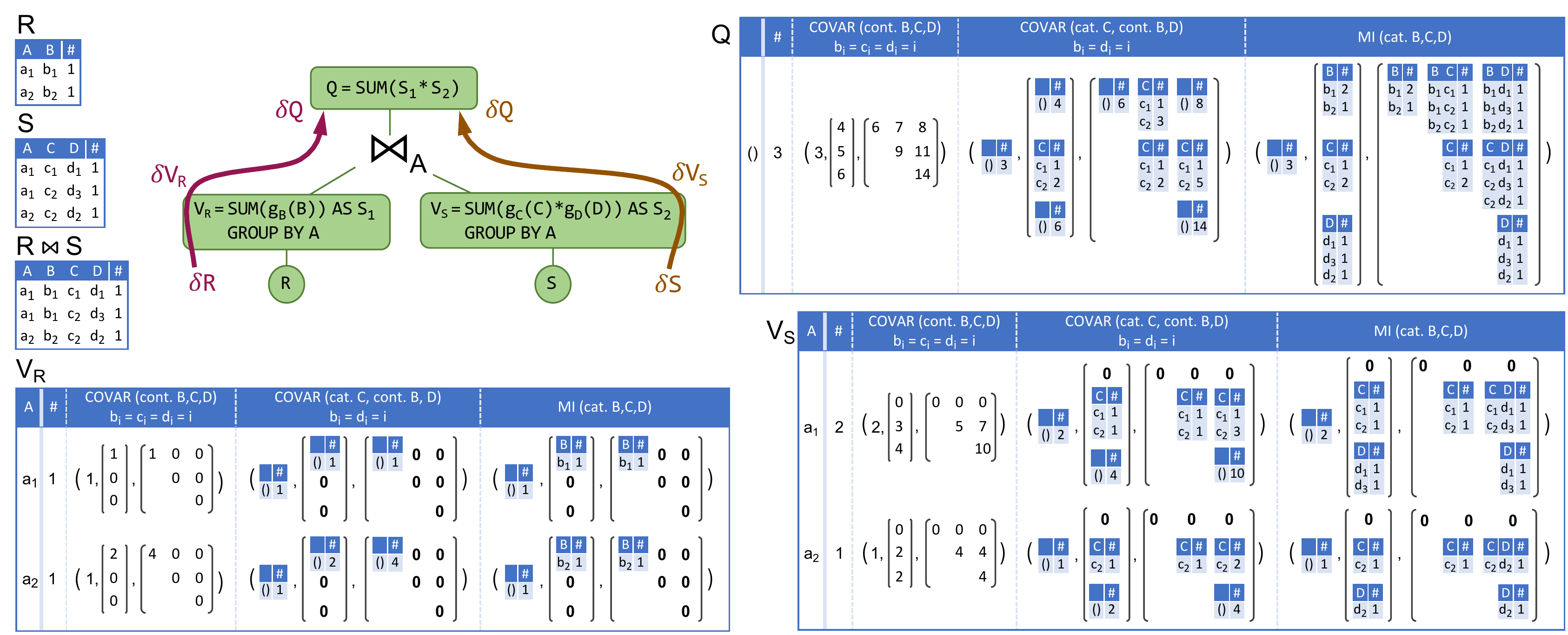}
\vspace{-1.8em}
\caption{Maintaining {\tt{SUM(g$_{\tt B}$(B)$\,*\,$g$_{\tt C}$(C)$\,*\,$g$_{\tt D}$(D))}} over the join $R(A,B)\Join S(A,C,D)$ in four
scenarios: tuple multiplicities using the $\mathbb{Z}$ ring (\#); COVAR matrix using the degree-3 ring (continuous B, C, D; also categorical C and continuous B, D); MI matrix using the degree-3 ring  (categorical B, C, D).
The symmetric matrix values are omitted.}
\label{fig:example}
\vspace{-0.8em}
\end{figure*}
\section{F-IVM By Example}

Consider the next query over relations $R(A,B)$ and $S(A,C,D)$:
\begin{lstlisting}[language=SQL, mathescape, columns=fullflexible, basicstyle=\small\ttfamily] 
        Q$\;\,$=$\;\,$SELECT$\;$SUM(g$_{\tt B}$(B)$\,*\,$g$_{\tt C}$(C)$\,*\,$g$_{\tt D}$(D))$\;$FROM$\;$R$\;$NATURAL$\;$JOIN$\;$S
\end{lstlisting}
The aggregates are from a ring $(\R,\Rplus,\Rprod,\Rzero,\Rone)$.
The {\tt SUM} operator uses the addition $\Rplus$ from $\R$. 
The attribute functions $\texttt{g}_{\tt B}$, $\texttt{g}_{\tt C}$, and $\texttt{g}_{\tt D}$ map attribute values to elements in $\R$. 

F-IVM exploits the distributivity of multiplication over the {\tt SUM} operator to push the aggregate past the join and later combine the partial aggregates to produce the query result. 
For instance, the view $V_S$ computes such partial sums over $S$:
\begin{lstlisting}[language=SQL, mathescape, columns=fullflexible, basicstyle=\small\ttfamily] 
        V$_\texttt{S}$ = SELECT A,$\;$SUM(g$_{\tt C}$(C)$\,*\,$g$_{\tt D}$(D))$\;$AS$\;$S$_\texttt{2}$ FROM S GROUP$\;$BY A
\end{lstlisting}
In the view $V_S$, we treat the $A$-values as keys and the aggregate $S_2$-values as payloads. 
Similarly, we can compute partial sums over $R$ as view $V_R$. These views are joined as depicted by the {\em view tree} in Figure~\ref{fig:example}, which is akin to a query plan. 

Figure~\ref{fig:example} gives a toy database consisting of relations $R(A,B)$ and $S(A,C,D)$, each mapping tuples to their multiplicity.
We next demonstrate how computing $Q$ over this database with different rings can support our application scenarios. 

{\bf Count Aggregate.}
We start with the simple count aggregate {\tt SUM(1)}. The aggregate values are from $\mathbb{Z}$, and $+$ and $*$ are the  arithmetic operations over $\mathbb{Z}$. The attribute functions $\texttt{g}_{\tt B}$, $\texttt{g}_{\tt C}$, and $\texttt{g}_{\tt D}$ map all values to 1.
Figure~\ref{fig:example} shows the contents of $V_R$ and $V_S$ on the toy database under the ring $\mathbb{Z}$ (the payload column \#).
To compute $Q$, we multiply the payloads of matching $A$-values from $V_R$ and $V_S$ and then sum up these products. 
The result of $Q$ is a relation mapping the empty tuple $()$ to the total number of tuples in the join of $R$ and $S$.

{\bf Linear Regression.}
Consider the problem of learning a linear function $f$ with parameters $\theta_{0}$, $\theta_{B}$ and $\theta_{C}$ that predicts the label $D$ given the features $B$ and $C$, where the training dataset is the natural join of our relations:
\begin{align*}
f(B, C) = \theta_{0} + \theta_{B} \cdot B + \theta_{C} \cdot C
\end{align*}
We assume that $B$, $C$, and $D$ have continuous domains; we consider the case when $C$ is a categorical attribute later on.

We can learn $f$ using batch gradient descent. 
This method iteratively updates the model parameters in the direction of the gradient to decrease the squared error loss and eventually converge to the optimal value.
The gradient of the square loss objective function requires the computation of three types of aggregates: 
the count aggregate \texttt{SUM(1)}, 
the linear aggregates \texttt{SUM(B)}, \texttt{SUM(C)}, and \texttt{SUM(D)}, and 
the quadratic aggregates \texttt{SUM(B$*$B)}, \texttt{SUM(B$*$C)}, \texttt{SUM(C$*$C)}, \texttt{SUM(B$*$D)}, and \texttt{SUM(C$*$D)}.
These aggregates suffice to capture the correlation between the features $B$ and $C$ and the label $D$~\cite{SOC:SIGMOD:16}.

F-IVM can compute all these aggregates using the query $Q$ and the same evaluation strategy from Figure~\ref{fig:example}!
The only needed adjustment is the replacement of the SQL \texttt{SUM} and $*$ operators with appropriate new sum and product operators. 

We treat this batch of aggregates as one compound aggregate $(c,\bm{s},\bm{Q})$, where $c$ is a scalar, $\bm{s}$ is a $3 \times 1$ vector with one sum of values per attribute, and $\bm{Q}$ is a $3 \times 3$ matrix of sums of products of values for any two attributes. This is the COVAR matrix.  
The compound aggregate can be pushed past the join similarly to the count aggregate discussed before. 
The payloads of keys carry these aggregates as values, see the payload column COVAR in Figure~\ref{fig:example}. 
The compound aggregates are from the {\em degree-3 matrix} ring $(\R, \Rplus^\R, \Rprod^\R, \Rzero, \Rone)$, 
where $\Rzero = (0, \bm{0}_{3\times1},\bm{0}_{3\times3})$, $\Rone = (1, \bm{0}_{3\times1},\bm{0}_{3\times3})$, and for $a = (c_a, \bm{s}_a, \bm{Q}_a) \in \R$ and $b = (c_b, \bm{s}_b, \bm{Q}_b)\in \R$:
\begin{align*}
a+^{\R}b &= (c_a + c_b, \bm{s}_a + \bm{s}_b, \bm{Q}_a + \bm{Q}_b) \\
a*^{\R}b &= (c_ac_b, c_b\bm{s}_a + c_a\bm{s}_b, c_b\bm{Q}_a + c_a\bm{Q}_b + \bm{s}_a\bm{s}_b^\texttt{T} + \bm{s}_b\bm{s}_a^\texttt{T})
\end{align*}
We use attribute names to index elements in $\bm{s}$ and $\bm{Q}$; for instance, $\bm{s} = [ \bm{s}_B \; \bm{s}_C \; \bm{s}_D]^\texttt{T}$. 
For each $X$-value $x$, where $X\in\{B,C,D\}$, the attribute function is ${\tt g}_{\tt X}(x) = (1, \bm{s}, \bm{Q})$, where $\bm{s}$ is a $3\times1$ vector with all zeros except $\bm{s}_{X} = x$, and $\bm{Q}$ is a $3 \times 3$ matrix with all zeros except $\bm{Q}_{XX} = x^2$.

In Figure~\ref{fig:example}, the payload $V_R(a_1) = {\tt g_B}(b_1)$ represents the mapped $B$-value $b_1$;
the payload $V_S(a_1) = {\tt g_C}(c_1) *^{\R} {\tt g_D}(d_1) +^{\R} {\tt g_C}(c_2) *^{\R} {\tt g_D}(d_2)$ represents the sum of products of the mapped $(C,D)$-pairs with the same $A$-value $a_1$; 
$V_R(a_2)$ and $V_S(a_2)$ are computed similarly. 
The result of $Q$ maps the empty tuple to the payload 
$V_R(a_1) *^{\R} V_S(a_1) +^\R V_R(a_2) *^{\R} V_S(a_2)$, yielding the count, the vector of aggregates {\tt SUM(X)}, and the matrix of aggregates {\tt SUM(X$*$Y)}, for $X,Y \in \{B,C,D\}$, over the join of $R$ and $S$.
Our approach significantly shares the computation across the aggregates: The scalar aggregates are used to scale up the linear and quadratic ones, while the linear aggregates are used to compute the quadratic ones.

{\bf Linear Regression with Categorical Attributes.}
Real-world datasets contain a mix of continuous and categorical attributes. 
The latter take on values from predefined sets of possible values (categories). 
It is common practice to one-hot encode categorical attributes as indicator vectors.

The COVAR matrix $\bm{Q}$ from above accounts for the interactions $\bm{Q}_{XY}$ = {\tt SUM(X$*$Y)} of attributes $X,Y \in\{B,C,D\}$ with a continuous domain.
Assume now that attribute $C$ is categorical and attributes $B$ and $D$ remain continuous. 
The interaction $\bm{Q}_{BC}$ captures the aggregates {\tt SUM(B)} per $C$-value:
\begin{lstlisting}[language=SQL, mathescape, columns=fullflexible, basicstyle=\small\ttfamily] 
        $\bm{Q}_{BC}$$\;$=$\;$SELECT$\;$C,$\;$SUM(B)$\;$FROM$\;$R$\;$NATURAL$\;$JOIN$\;$S$\;$GROUP$\;$BY$\;$C
\end{lstlisting}
Using the group-by clause ensures a compact representation of one-hot encoded $C$-values and that $\bm{Q}_{BC}$ considers only the $C$-values that exist in the join result.

We unify the representation of aggregates for continuous and categorical attributes by composing the degree-$3$ matrix ring with the ring over relations~\cite{Nikolic:SIGMOD:18} as follows:
we use relations as values in $c$, $\bm{s}$, and $\bm{Q}$ instead of scalars;
we use union and join instead of scalar addition and multiplication;
we use the empty relation $\bm{0}$ as zero.
The operations $+^\R$ and $*^\R$ over triples $(c, \bm{s}, \bm{Q})$ remain unchanged.

The attribute function ${\tt g}_{\tt X}$ now depends on whether $X$ is continuous or categorical.
For any $X$-value $x$, 
${\tt g}_{\tt X}(x) = (\bm{1}, \bm{s}, \bm{Q})$, 
where $\bm{1} = \{() \mapsto 1\}$,
$\bm{s}$ is a $3\times1$ vector with all $\bm{0}s$ 
except $\bm{s}_X = \{ () \mapsto x \}$ if $X$ is continuous and $\bm{s}_X =\{ x \mapsto 1 \}$ otherwise, 
and $\bm{Q}$ is a $3 \times 3$ matrix with all $\bm{0}s$ 
except $\bm{Q}_{XX} = \{ () \mapsto x^2 \}$ if $X$ is continuous and $\bm{Q}_{XX} = \{ x \mapsto 1 \}$ otherwise.


Figure~\ref{fig:example} shows the contents of $V_R$, $V_S$, and $Q$ when using the generalized degree-3 matrix ring with relational values (see the payload column COVAR with $C$ as categorical).
The payload in $Q$ captures the count aggregate $c$, 
the vector $\bm{s}$ of aggregates $\bm{s}_B$ = {\tt SUM(B)}, $\bm{s}_C$ = {\tt SUM(1)} grouped by $C$, and $\bm{s}_D$ = {\tt SUM(D)}, and the matrix $\bm{Q}$ of aggregates including $\bm{Q}_{BC}$ = {\tt SUM(B)} grouped by $C$ and $\bm{Q}_{BD}$ = {\tt SUM(B$*$D)}.
The computation follows the same pattern as with the count aggregate and linear regression with continuous attributes. The only difference is due to the ring used for payloads.

{\bf Mutual Information (MI).}
The MI of two discrete random variable $X$ and $Y$ is defined as:
\begin{align*}
I(X,Y) = \sum_{x \in Dom(X)}\sum_{y \in Dom(Y)} p_{XY}(x,y) \log\left(\frac{p_{XY}(x,y)}{p_X(x)\,p_{Y}(y)}\right)
\end{align*}
where $p_{XY}$ is the joint probability mass function of $X$ and $Y$, and $p_X$ and $p_Y$ are the probability mass functions of $X$ and respectively $Y$. 
In our database setting, we can capture the joint distribution of two categorical attributes $X$ and $Y$ and the two marginal distributions using four count aggregates:
$C_{\emptyset}$ = {\tt SUM(1)}, $C_{X}$ = {\tt SUM(1)} grouped by $X$, $C_{Y}$ = {\tt SUM(1)} grouped by $Y$, and $C_{XY}$ = {\tt SUM(1)} grouped by $(X,Y)$.
The MI of $X$ and $Y$ is then computed as:
\begin{align*}
I(X,Y) = \sum_{x \in Dom(X)}\sum_{y \in Dom(Y)} \frac{C_{XY}(x,y)}{C_{\emptyset}} \log{\frac{C_{\emptyset}\,C_{XY}(x,y)}{C_X(x)\,C_Y(y)}}
\end{align*}

We compute the MI for all pairs $(X,Y)$ of categorical attributes. The aggregates $C_\emptyset$, $C_X$, and $C_{XY}$ are exactly those computed for the COVAR matrix over categorical attributes. 
We can thus assemble the $C_X$ aggregates into a vector and the $C_{XY}$ aggregates into a matrix, and share their computation as in the linear regression case. When computing the MI for continuous attributes, we first discretize their values into bins of finite size and then follow the same steps as with computing the MI for categorical attributes.

The views $V_R$, $V_S$, and $Q$ from Figure~\ref{fig:example} capture the aggregates $C_\emptyset$, $C_X$, and $C_{XY}$ of categorical attributes $X,Y\in\{B,C,D\}$ using the degree-3 matrix ring with relational values (the last payload column MI). 
The payload in $Q$ consists of the count aggregate $C_\emptyset$, the vector $\bm{s}$ of {\tt SUM(1)} aggregates grouped by $X$, and the matrix $\bm{Q}$ of {\tt SUM(1)} aggregates grouped by $(X,Y)$, for $X,Y\in\{B,C,D\}$. As in the previous examples, the computation over keys remains the same.

The MI of two attributes quantifies their degree of correlation~\cite{murphy2013}: A value close to 0 means they are almost independent, while a large value means they are highly correlated. 
It can identify attributes that predict (are highly correlated with) a given label attribute and can thus be used for model selection~\cite{murphy2013}. It can also be used for learning the structure of Bayesian networks. The Chow-Liu algorithm~\cite{Chow-Liu-trees:1968} constructs an optimal tree-shaped Bayesian network with one node for each attribute. It proceeds in rounds and in each round it adds an edge between two nodes such that their pairwise MI is maximal among all pairs of attributes not chosen yet.

\begin{figure*}[t]
  \begin{subfigure}{.245\textwidth}
    \centering
    \includegraphics[width=1.0\linewidth]{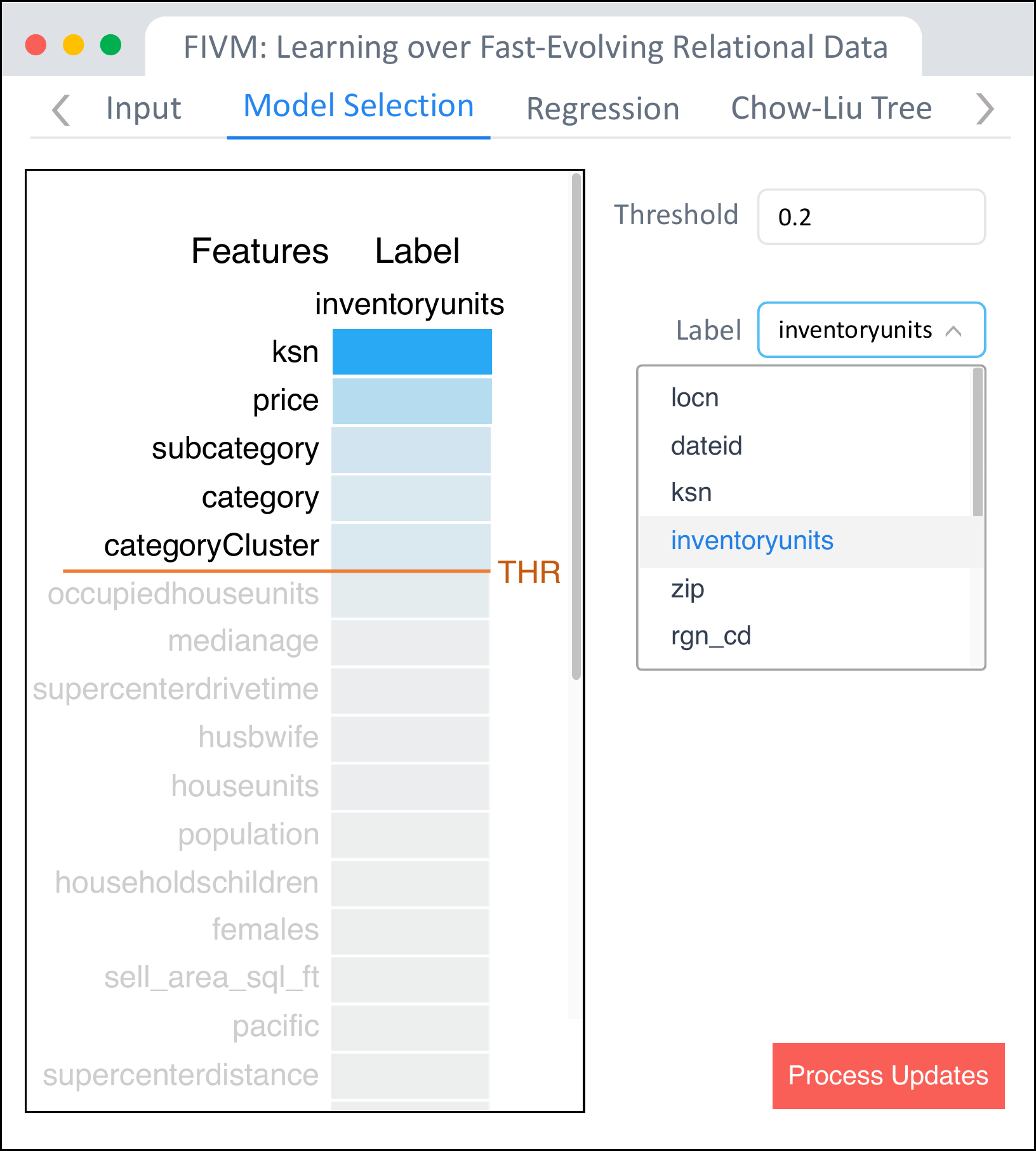}
    \caption{Model Selection}
    \label{fig:sfig1}
  \end{subfigure}
  \begin{subfigure}{.245\textwidth}
    \centering
    \includegraphics[width=1.0\linewidth]{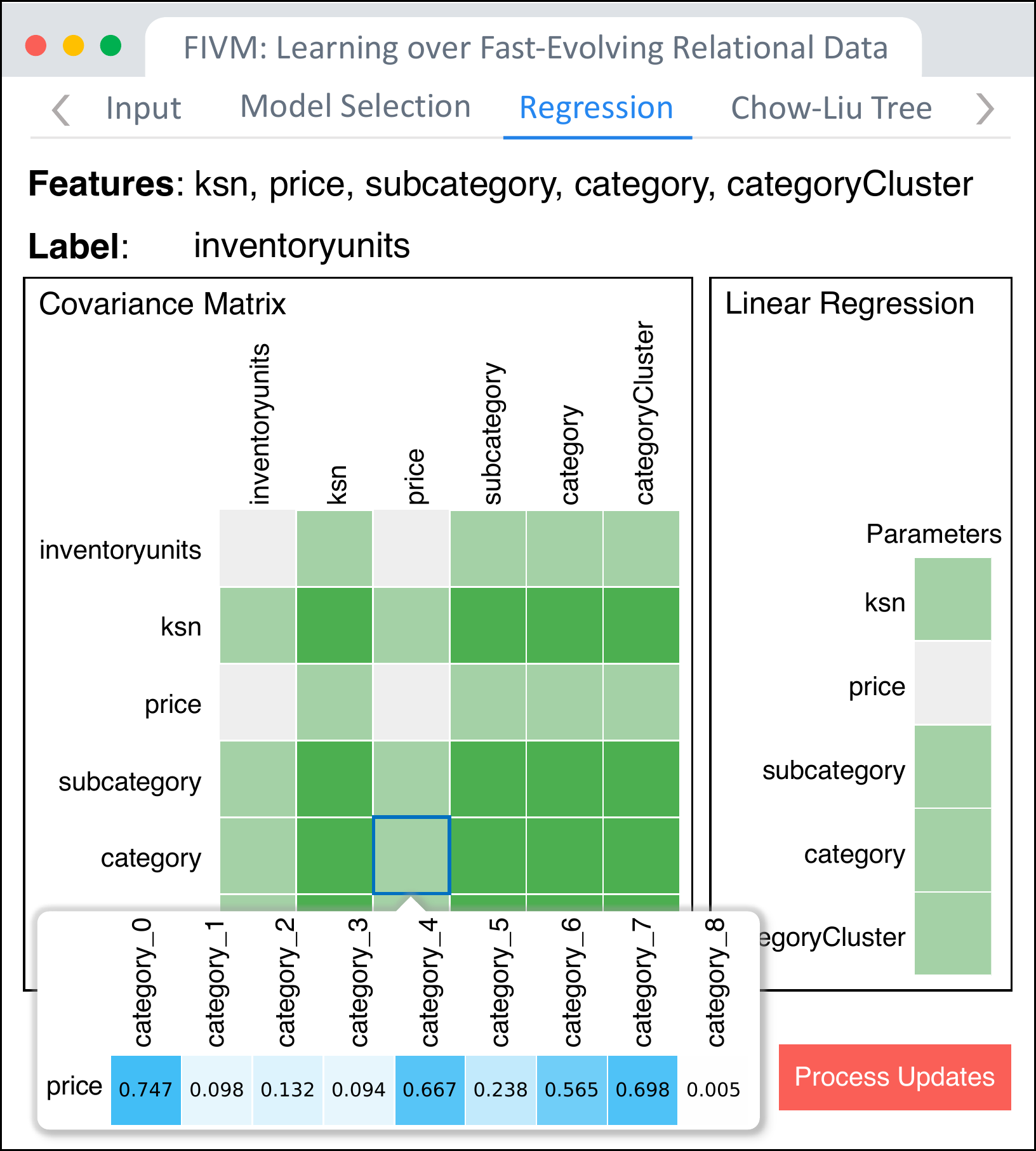}
    \caption{Regression}
    \label{fig:sfig2}
  \end{subfigure}
  \begin{subfigure}{.245\textwidth}
    \centering
    \includegraphics[width=1.0\linewidth]{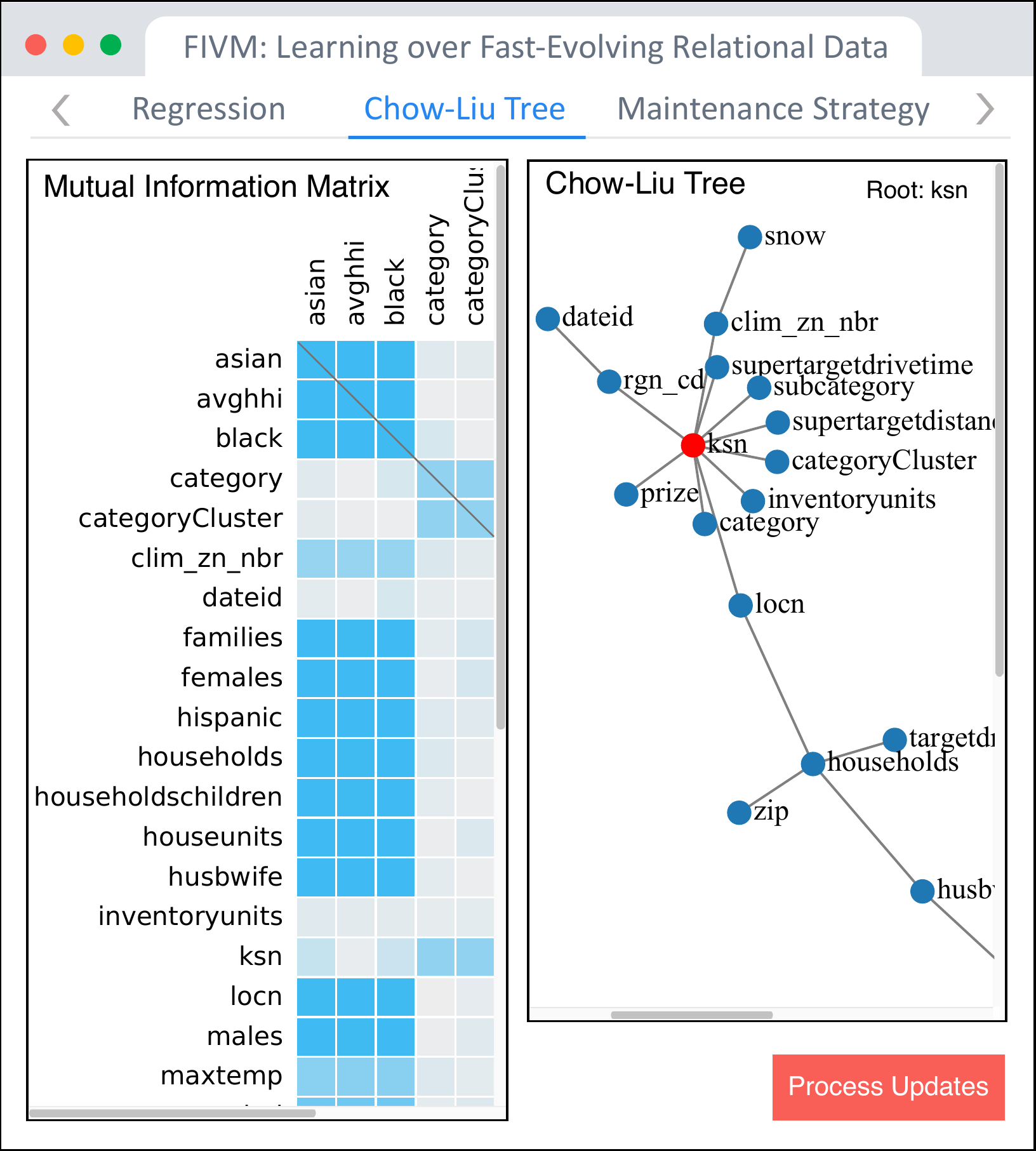}
    \caption{Chow-Liu tree}
    \label{fig:sfig3}
  \end{subfigure}
  \begin{subfigure}{.245\textwidth}
    \centering 
    \includegraphics[width=1.0\linewidth]{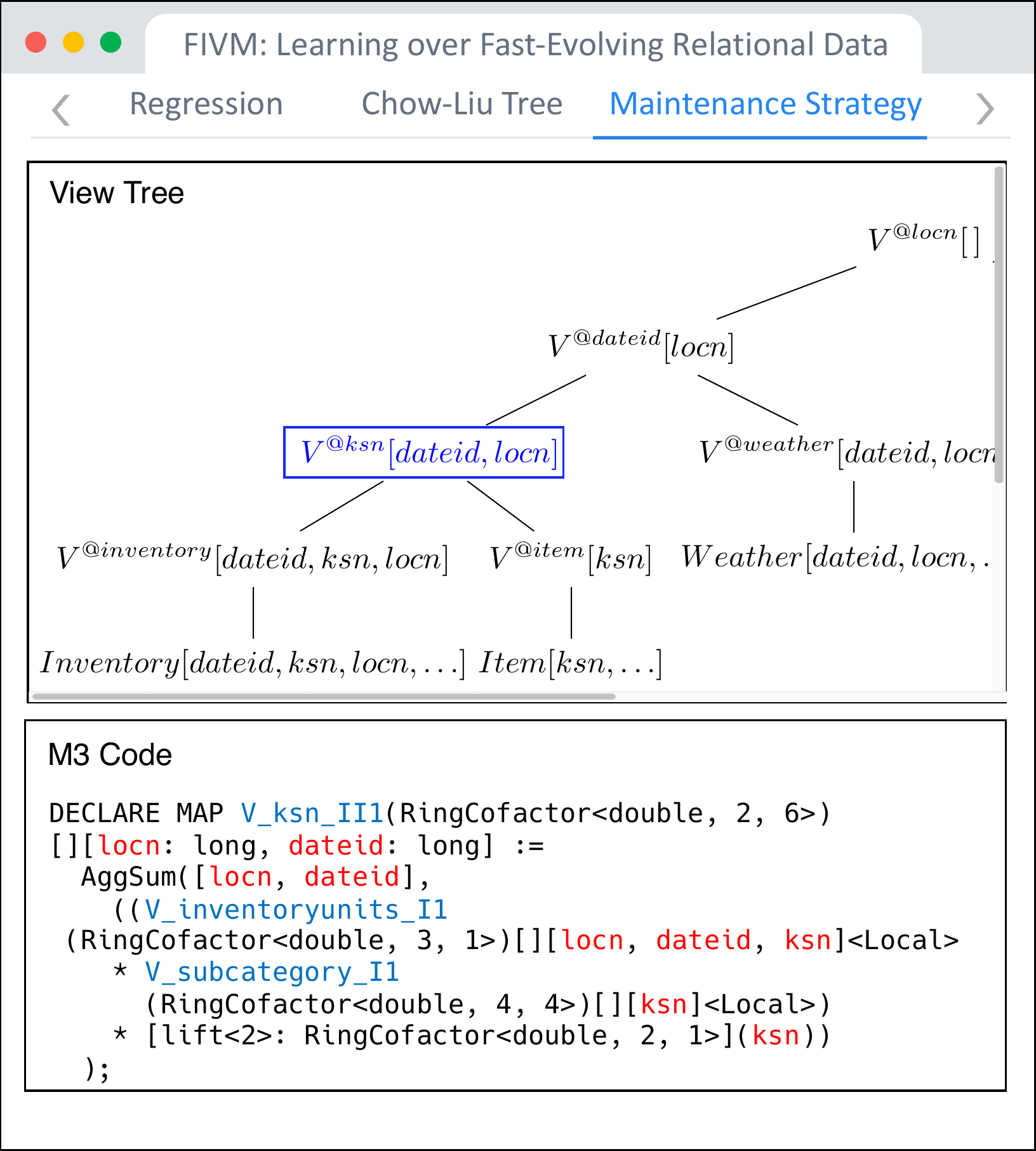}
    \caption{Maintenance Strategy}
    \label{fig:sfig4} 
  \end{subfigure}
  \vspace{-0.4em}
  \caption{F-IVM's web user interface: (a) model selection using pairwise mutual information of attributes with a given label; inspect (b) learning the ridge linear regression model with the selected features and label; (c) the mutual information matrix and the Chow-Liu tree; (d) the F-IVM view tree and M3 code for views. }
  \vspace{-0.4em}
  \label{fig:snapshots}
\end{figure*}

{\bf Incremental Maintenance.}
Figure~\ref{fig:example} shows the leaf-to-root path taken to maintain the query result under updates $\delta{R}$ to $R$. The delta $\delta{V_R}$ captures the change in $V_R$:
\begin{lstlisting}[language=SQL, mathescape, columns=fullflexible, basicstyle=\small\ttfamily] 
        $\delta$V$_\texttt{R}$ = SELECT A,$\;$SUM(g$_{\tt B}$(B))$\;$AS$\;$S$_\texttt{1}$ FROM $\delta$R GROUP$\;$BY A
\end{lstlisting}
The delta $\delta{V_R}$ further joins with $V_S$ to compute $\delta{Q}$.

The update $\delta{R}$ may consist of both inserts and deletes, which are encoded as keys with positive and respectively negative payloads. In our examples, a negative payload is $-1$ for the count aggregate and $(\{()\mapsto -1\}, \bm{0}_{3\times1}, \bm{0}_{3\times3})$ for the compound aggregate with relational values, where $\bm{0}_{m\times n}$ is the $m \times n$ matrix whose entries are the empty relation.

\section{User Interaction}

Figure~\ref{fig:snapshots} depicts snapshots of F-IVM's user interface.

In the {\it Input} tab (not shown), the user chooses the database and gives the query defining the initial training dataset. A sequence of updates is prepared for each database. The MI and COVAR matrices and the  applications built on top of them are first computed over this initial training dataset.

The {\it Model Selection} tab allows the user to specify a label attribute for a predictive model and an MI threshold. It then depicts the list of all attributes ranked based on their pairwise MI with the label. Only the attributes above the threshold are selected as features of the model. 
\nop{The interface uses darker/lighter blue to depict larger/smaller MI values (the values are not shown). 
To show the effect of the updates to the human eye, }
F-IVM processes one bulk of 10K updates before pausing for one second. The users can then observe how relevant attributes become irrelevant to predicting the label or vice-versa.

The {\it Regression} tab allows users to inspect the ridge linear regression model with the features and label chosen in the previous tab. F-IVM updates the COVAR matrix after each bulk of updates. Then, a batch gradient descent solver resumes the convergence of the model parameters using gradients that are made of the previous parameter values and the new COVAR matrix~\cite{Nikolic:SIGMOD:18}.
A dark green or white cell in this matrix is a 2D or respectively 0D tensor representing the interaction between two categorical or respectively two continuous attributes. A white cell is a 1D tensor that represents the interaction between a categorical and a continuous attribute.
 The same color coding is used for the model vector. A tensor is shown when clicking its cell.

The {\it Chow-Liu Tree} tab depicts the MI matrix for all pairs of attributes in the training dataset and the Chow-Liu tree constructed using this matrix. After each bulk of 10k updates, F-IVM updates the matrix and the tree and then pauses.

\nop{It is implemented on top of the DBToaster backend~\cite{DBT:VLDBJ:2014}, which provides state-of-the-art code optimization and compilation for incremental maintenance of relational queries over databases. F-IVM replaces DBToaster's frontend with a new frontend that supports efficient maintenance strategies for batches of aggregates instead of individual aggregates. Furthermore,  F-IVM provides novel rings to capture expressive dynamic computation necessary for the maintenance of common relational learning applications. Using rings, both inserts and deletes are treated alike by F-IVM. Also, by changing the meaning of the sum and product ring operators as well as of the ring support, the same symbolic computation expressed using sums and products over elements from the ring support may capture the behavior of different applications.
}

The {\it Maintenance Strategy} tab depicts the F-IVM view tree for the input query. For each view it shows the code for this view in the M3 interpreted representation language~\cite{DBT:VLDBJ:2014}.
\nop{, which is a layer between SQL and C++
}
We use the DBToaster backend~\cite{DBT:VLDBJ:2014} to compile M3 code into efficient C++ code before running it on the updates.

\bibliographystyle{abbrv}
\bibliography{bibliography}

\end{document}